\documentclass[sn-mathphys-num]{sn-jnl}

\usepackage{graphicx}%
\usepackage{multirow}%
\usepackage{amsmath,amssymb,amsfonts}%
\usepackage{amsthm}%
\usepackage{mathrsfs}%
\usepackage[title]{appendix}%
\usepackage{xcolor}%
\usepackage{textcomp}%
\usepackage{manyfoot}%
\usepackage{booktabs}%
\usepackage{algorithm}%
\usepackage{algorithmicx}%
\usepackage{algpseudocode}%
\usepackage{listings}%

\theoremstyle{thmstyleone}%
\theoremstyle{thmstyletwo}%

\theoremstyle{thmstylethree}%

\begin{document}

\title[Article Title]{Undulator Radiation from a Single Electron:\\
A Temporal Double-Slit Experiment}


\author*[1]{\fnm{Shaukat} \sur{Khan}}\email{shaukat.khan@tu-dortmund.de}

\author*[2,4]{\fnm{Yuya} \sur{Asai}}\email{asaiyuya0705@hiroshima-u.ac.jp}

\author[1]{\fnm{Zohair} \sur{Usfoor}}\email{zohair.usfoor@tu-dortmund.de}

\author[3,4]{\fnm{Tatsuo} \sur{Kaneyasu}}\email{kaneyasu@ims.ac.jp}

\author[1,8]{\fnm{Carsten} \sur{Mai}}\email{carsten.mai@tu-dortmund.de}

\author[3,5,6]{\fnm{Hiroshi} \sur{Miyauchi}}\email{hiroshi.miyauchi@kek.jp}

\author[4]{\fnm{Yasuaki} \sur{Okano}}\email{yokano@ims.ac.jp}

\author[1]{\fnm{Arjun} \sur{Radha Krishnan}}\email{arjun.krishnan@tu-dortmund.de}

\author[7]{\fnm{Wa'el} \sur{Salah}}\email{wsalah@hu.edu.jo}

\author[3,5,6]{\fnm{Miho} \sur{Shimada}}\email{miho.shimada@kek.jp}

\author[1]{\fnm{Vivek} \sur{Vijayan}}\email{vivek.vijayan@tu-dortmund.de}

\author[2,4,5]{\fnm{Masahiro} \sur{Katoh}}\email{mkatoh@ims.ac.jp}

\affil[1]{\orgdiv{Center for Synchrotron Radiation (DELTA) and Department of Physics},\\ \orgname{TU Dortmund University}, \orgaddress{\street{Maria-Goeppert-Mayer-Str.\ 2},\\ \city{44227 Dortmund}, \country{Germany}}}

\affil[2]{\orgdiv{Graduate School of Advanced Science and Engineering}, \orgname{Hiroshima University}, \orgaddress{\street{1-3-1 Kagamiyama}, \city{Higashi-Hiroshima}, \postcode{739-8526}, \country{Japan}}}

\affil[3]{\orgdiv{Graduate School of Advanced Studies}, \orgname{SOKENDAI}, \orgaddress{\street{Shonan Village, Hayama}, \city{Kanagawa}, \postcode{210-0193}, \country{Japan}}}

\affil[4]{\orgdiv{UVSOR Synchrotron Facility}, \orgname{Institute for Molecular Science, National Institutes of Natural Sciences}, \orgaddress{\street{38 Nishigo-Naka}, \city{Okazaki}, \postcode{444-8585}, \country{Japan}}}

\affil[5]{\orgdiv{Research Institute for Synchrotron Radiation Science (HiSOR)}, \orgname{Hiroshima University}, \orgaddress{\street{2-313 Kagamiyama}, \city{Higashi-Hiroshima}, \postcode{739-0046}, \country{Japan}}}

\affil[6]{\orgdiv{High Energy Accelerator Research Organization}, \orgname{KEK}, \orgaddress{\street{1-1 Oho}, \city{Tsukuba}, \postcode{305-0801}, \country{Japan}}}

\affil[7]{\orgdiv{Department of Physics}, \orgname{The Hashemite University}, \orgaddress{\street{453J+5C5 Damascus Hwy}, \city{Zarqa}, \postcode{13133}, \country{Jordan}}}

\affil[8]{\orgdiv{Helmholtz-Zentrum Berlin (HZB)}, \orgaddress{\street{Albert-Einstein-Str.\ 15}, \city{12489 Berlin}, \country{Germany}}}


\abstract{Double-slit diffraction studies with photons or massive particles rank among the
most beautiful experiments in physics.
In particular, measurements at very low intensities demonstrate the particle-wave
duality and the coherent superposition of states very clearly.
In this paper, 
low-intensity double-slit experiments in the time domain are presented
measuring the spectral distribution of synchrotron light from a single relativistic electron 
in a storage ring.
In two consecutive radiation sources (so-called undulators) with a magnetic detour between them,
electrons emit two temporally separated light pulses leading to a spectrum
with interference fringes -- very much like the angular distribution of light
behind two spatially separated slits.
Independent experiments at two synchrotron light sources
(DELTA in Germany and UVSOR-III in Japan) 
directly demonstrate that the spectral distribution of accumulated 
synchrotron light from a single electron is essentially 
the same as the spectrum from a beam of many electrons.
While the latter is usually explained as interference between 
electromagnetic waves from the two undulators, 
the single-electron experiments demonstrate that 
coherent photon emission is delocalized over several meters
and the accumulated spectral distribution
exhibits a deterministic interference pattern at small wavelengths.
The experiments presented here were conducted with near-ultraviolet light 
to avoid an elaborate in-vacuum setup, 
but the very wide spectral range of synchrotron radiation, from infrared light to X-rays, enables access to regimes not available in laser-based quantum optics experiments.}




\maketitle

\section{Introduction}\label{sec1}

Synchrotron radiation from two consecutive sources emitted
by a single relativistic electron in a storage ring is a time-domain
analogy of a double-slit experiment with feeble light.
The measurements described below were conducted independently at two synchrotron light sources.
In contrast to other setups in the rapidly developing field of quantum science and technology, 
synchrotron radiation can access a wide range of wavelengths, 
from millimeters to the sub-nanometer regime, 
while the devices controlling the photon emission are magnets of macroscopic dimensions.

The classical double-slit experiment, first performed by T. Young in 1801, demonstrated 
``some strong resemblance between the nature of sound and that of light'' \cite{Young1804}, namely their wave-like behavior. 
Monochromatic light waves passing through two slits give rise to a pattern of constructive and destructive interference on a screen. 
Furthermore, double-slit experiments have illustrated some perplexing features of quantum mechanics.
Interference patterns from de Broglie waves of massive particles like
electrons \cite{Joensson1961}, neutrons \cite{Zeilinger1988}, 
and even C$_{70}$ molecules \cite{Brezger2002} have proven the fundamental concept of
particle-wave duality. 
Another key finding was that interference also occurs when the light source is so weak that only one photon is emitted at a time \cite{Taylor1909}.
For the double-slit case, the pattern emerging from a large number of individual photons results from a coherent superposition of the probability amplitudes for the indistinguishable pathways through one slit or the other.

The angular distribution in a double-slit experiment exhibits intensity maxima with a spacing given by the slit separation and an
envelope corresponding to the diffraction pattern from one slit.
In the time domain,
the spectrum of two consecutive and nearly monochromatic light pulses has a similar appearance.
The positions of the maxima in the spectrum depend on the temporal pulse separation,
and the envelope is given by the time structure of each pulse.
In the case of synchrotron radiation from an undulator, 
a sequence of magnets with alternating polarity \cite{Wiedemann2010},
the pulse has a rectangular temporal shape, i.e., a constant intensity between two limits, 
which is analogous to a lateral slit with two edges.

\begin{figure}[t]
\centering
\includegraphics[width=0.9\textwidth]{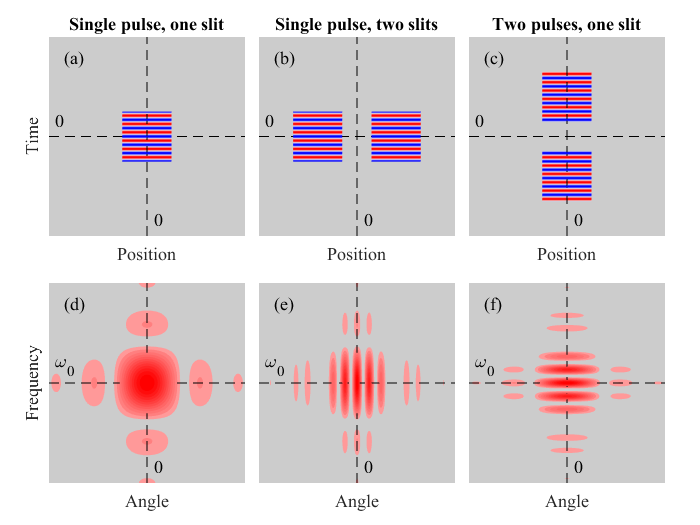}
\caption{
Top: Schematic of the electric field of light pulses as function of transverse position and time. 
(a) A single pulse with rectangular temporal profile after passing a slit, 
(b) a single pulse after two slits, 
(c) two consecutive pulses after a single slit.
Bottom:
Distribution of light intensity as function of angle and frequency in the 
far field with panels (d-f) corresponding to (a-c).}
\label{fig1}
\end{figure}

Figure 1 (a) represents the electric field of a single light pulse with a 
rectangular envelope in lateral position as well as in time.
The angular distribution and the spectrum sketched in Fig. 1 (d) are given by the square of a sinc function, i.e., both distributions are proportional to 
$\left\{ \sin \left[ u \right]/u \right\}^2$ with $u$ being angle or frequency. 
Two pulses with lateral and temporal separation are shown in Figs. 1 (b) and (c), respectively.
Consequently, an interference pattern appears in the angular (e) 
or spectral distribution (f).
It is thus justified to refer to the latter case as a ``temporal double-slit experiment''.
However, the analogy is not perfect.
While the light pulse is uniform within each slit and the
angular pattern is centered at zero,
its oscillatory temporal structure results 
in a spectrum centered about frequency $\omega_0$. 

The temporal double-slit experiments presented in this paper involve radiation from two undulators, each comprising $N_{\rm U}$ magnetic periods, i.e., $2 N_{\rm U}$ dipole 
magnets with alternating polarity.
The magnetic field is given by $B(s) = B_0 \sin (2 \pi s/\lambda_{\rm U})$, 
where $s$ is the coordinate along the undulator and $\lambda_{\rm U}$ is the period length.
When relativistic electrons with kinetic energy $E_0$ pass an undulator on a sinusoidal trajectory, 
they emit synchrotron radiation with a central wavelength of \cite{Wiedemann2010}
\begin{equation}
\lambda_0 = \frac{\lambda_{\rm U}}{2 \gamma^2} 
\left( 1 + \frac{K^2}{2} + \gamma^2 \theta^2 \right).
\end{equation}
Here, 
$\gamma = E_0/(m_e c^2)$ is the Lorentz factor with the electron rest mass $m_e$
and the speed of light $c$,
$K = 93.4 \, \lambda_{\rm U} \left[ {\rm m} \right]  B_0 \left[ {\rm T} \right]$
is the undulator strength parameter, and $\theta$ is the observation angle
with respect to the undulator axis.
With $\gamma^2$ of the order of $10^6$ to $10^7$ and a 
period length on the centimeter scale, the radiation wavelength may range from visible 
light ($K > 10$) to the X-ray regime ($K < 1$).
Since the electric field of undulator radiation has a rectangular temporal envelope
over $N_{\rm U}$ optical cycles, the on-axis power spectrum ($\theta = 0$)
follows a squared sinc function 
\begin{equation}
P_{\rm U}(\lambda) \propto \frac{\lambda^4}{\gamma^2}
\left\{ 
\frac{\sin \left[ \pi N_{\rm U} 
\left( \lambda_0/\lambda -1 \right) \right]}
{\pi N_{\rm U} \left( \lambda_0/\lambda -1 \right) } 
\right\}^2
\end{equation}
with a width proportional to $1/N_{\rm U}$.

A configuration of two undulators tuned to the same wavelength $\lambda_0$ with a so-called chicane between them is known as ``optical klystron'' 
in the context of free-electron lasers \cite{Elleaume1983,Kornyukhin1985}.
A magnetic chicane, typically formed by four dipole magnets,
is characterized by the longitudinal dispersion parameter 
$R_{56} \equiv \Delta z / (\Delta E/E_0)$, 
which describes the path length difference $\Delta z$ 
caused by an energy deviation $\Delta E$.
Compared to a straight passage, the electron path 
is lengthened by $R_{56}/2$, which corresponds to the delay between the pulses
emitted in each undulator.  
The on-axis spectrum of an optical klystron is given by 
\begin{equation}
P_{\rm OK}(\lambda) = 2 P_{\rm U}(\lambda) \left\{
1 + f \cos \left[ 2\pi \left( N_{\rm U} + N_{\rm D} \right)
\frac{\lambda_0}{\lambda }  
\right]
\right\},
\end{equation}   
where the delay is expressed as number of optical cycles
$N_{\rm D} = R_{56}/(2 \lambda )$.
The electron energy spread $\sigma_{\Delta E/E}$ reduces the
visibility of the interference fringes according to 
$f\equiv \exp \left\{ -\frac{1}{2} 
\left[ 4 \pi \sigma_{\Delta E/E} \left( N_{\rm U} + N_{\rm D} \right)
\right]^2
\right\} < 1$.

These and other properties of undulator radiation are readily obtained in the frame
of classical electrodynamics by integrating over Li\'{e}nard-Wiechert
potentials generated from a charge along its path 
\cite{Jackson1999,Wiedemann2010,Hofmann2007}.
The spectrum is given by interference between radiation from successive undulator periods for an electron bunch producing a quasi-continuous electric field.
However, a single electron randomly emits only one photon after passing an undulator
about 100 times,
and it is not self-evident whether the emission takes place at a particular
magnetic pole or is delocalized over the whole undulator.

A quantized radiation field can be described by an ensemble of uncoupled harmonic oscillators,
where the eigenstates of the annihilation operator are best suited
to describe a classical light wave \cite{Fox2006,Ou2017,Bachor2019}.
These so-called coherent states approximately represent an oscillating electric field with well-defined amplitude and phase.
A radiation field is generally composed of a coherent or incoherent mixture of pure states. 
Its nature reveals itself when studying statistical properties such as fluctuations of intensity or photon arrival time, 
or by observing correlation functions or interference patterns.
Measurements have verified that radiation from an undulator is statistically consistent
with a coherent state \cite{Lobach2020}, 
and for photon energies small compared to the electron energy,
the accumulated radiation from a single electron should have the same properties 
as conventional synchrotron light. 

Experiments with single electrons were conducted since the early days of storage rings.
In the 1960s, visitors at AdA, the world's first $e^+e^-$ collider in Frascati, Italy, 
were shown synchrotron light from one electron, which was visible to the naked eye \cite{Bonolis2018}.
The stochastic nature of radiation from a single electron was studied at
VEPP-2 and VEPP-3 in Novosibirsk, Russia \cite{Pinayev1994,Aleshaev1995,Pinayev1996}.
Single electrons were employed for metrological purposes at BESSY, BESSY II, 
and MLS \cite{Riehle1988,Brandt2007,Klein2010} in Berlin, Germany, 
and recent studies were carried out at the IOTA ring at Fermilab, USA 
\cite{Lobach2020,Romanov2021,Lobach2022,Stancari2024}.

\section{Results}

Experimental investigations of single-electron spectra from undulators 
were undertaken independently at two synchrotron light sources, 
one being the 1.5-GeV electron storage ring DELTA in Dortmund, Germany \cite{Khan2024}, 
the other UVSOR-III in Okazaki, Japan, with a beam energy of 750\,MeV \cite{Asai2025}.
In both cases, a beam of low current was injected into the storage ring and then reduced to a single electron by moving an obstacle close to the beam, a so-called scraper, which shortens the beam lifetime from hours to minutes.
Observing synchrotron radiation with a PMT at a sufficiently low number of circulating electrons, 
the loss of a single particle was detected as a discrete step in the PMT signal
-- see Fig.\ 5 in the Methods section.
The scraper was then retracted when only one electron was left.
Since a PMT is not able to measure the photon energy, different techniques were applied to obtain spectral information on single-electron undulator radiation.

\begin{figure}[t]
\centering
\includegraphics[width=0.9\textwidth]{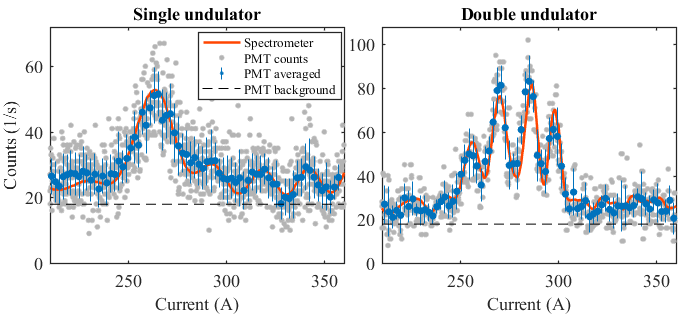}
\caption{Intensity of synchrotron light after passing a 400-nm bandpass filter while scanning the coil current of a single electromagnetic undulator (left) and
two consecutive undulators (right) at the DELTA storage ring.
Gray: PMT count rate from a single electron. 
Blue: Average PMT count rate and standard deviation within current intervals of 2\,A.
Dashed line: Average PMT background rate measured separately.
Red: Spectrometer data with 2\,mA electron current normalized to the PMT counts above the background. 
}
\label{fig2}
\end{figure}

\begin{figure}[h]
\centering
\includegraphics[width=0.92\textwidth]{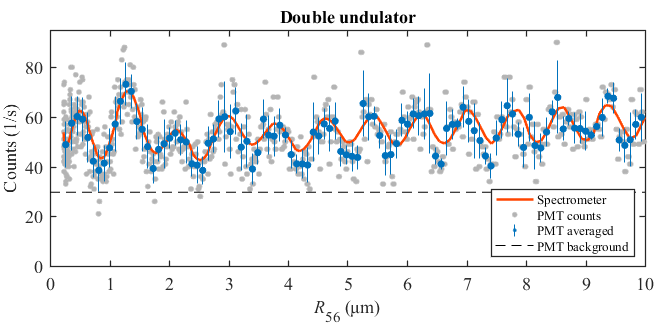}
\caption{Intensity of synchrotron light after passing a 400-nm bandpass filter while scanning the 
longitudinal dispersion $R_{56}$ of a magnetic chicane between two consecutive undulators at the DELTA storage ring.
Gray: PMT count rate from a single electron. 
Blue: Average PMT count rate and standard deviation within $R_{56}$ 
intervals of 0.1\,\textmu m.
Dashed line: Average PMT background rate measured separately.
Red: Spectrometer data with 2\,mA electron current normalized to the PMT counts above the background.}
\label{fig3}
\end{figure}

\begin{figure}[t]
\centering
\includegraphics[width=0.9\textwidth]{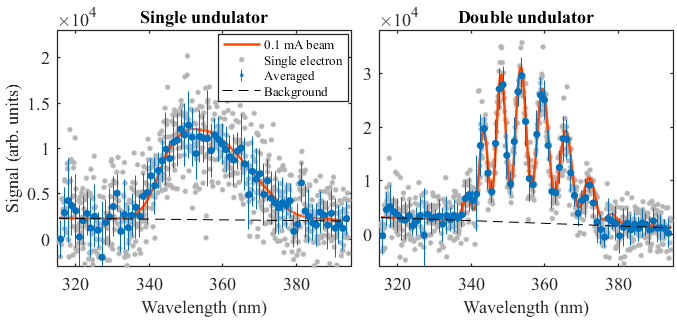}
\caption{Spectra of a single undulator (left) and
two consecutive undulators (right) at UVSOR-III.
Gray: Single-electron spectrometer data averaged over vertical CCD pixels as function of wavelength after subtracting a dark image.  
Blue: Average and standard deviation over wavelength intervals of 1\,nm for a single electron.
Dashed line: Interpolated residual background.
Red: Spectrometer data for a 0.1-mA beam normalized to the single-electron spectra above the background.}
\label{fig4}
\end{figure}

At DELTA, undulator radiation passing an aperture (acceptance $\pm 0.4$\,mrad) and a bandpass filter 
(wavelength 400\,nm $\pm $ 5\,nm) was monitored by a PMT
while the central wavelength $\lambda_0$ was scanned
by changing the current of the electromagnetic poles.
This is not exactly a measurement of the undulator spectrum, since the
spectral distribution slightly changes during the scan. 
It is, however, a convenient method to compare spectral information obtained from a few-mA beam and a single electron.
At a beam current of about 2\,mA ($4.8\cdot 10^9$ electrons), the radiation was coupled into a fiber spectrometer equipped with a grating and a CCD array.
Recording one spectrum per second, the whole scan took a few minutes.
Photons from a single electron were counted after the same aperture using a PMT while
varying the central wavelength over typically 45 minutes.
These scans were performed for a single undulator and 
for two versions of an optical klystron, one with the second undulator directly following the chicane, the other with a 1\,m long field-free drift space preceding the second undulator.
The drift space was introduced to check for any decoherence effect 
in the photon emission.
However, none of the experimental results show any significant discrepancy between the 
signal from a few-mA beam and a single electron.
Examples are given in Fig. 2 and further details on the measurements are
discussed in the Methods section.

In another experiment, the magnetic field of both optical-klystron undulators was tuned to a central wavelength of 400\,nm and the light passing the 400-nm 
bandpass filter was recorded while the current of the chicane magnets was steadily increased.
As before, radiation from a few-mA beam was recorded by a fiber spectrometer within minutes 
while photons from a single electron were counted using a PMT,
where the slowest scan took two hours.
The result for many electrons (red curve in Fig.\ 3)
is again consistent with the single-electron data (gray and blue dots).
In this particular magnetic setup, a chicane current from 42\,A to 136\,A corresponds to a
longitudinal dispersion of $R_{56} \approx 2$\,\textmu m to 10\,\textmu m as
shown in the figure.
Spectrometer and PMT data over the full scan from zero up to 450\,A
are presented in the section on Methods (Fig. 7).
 
At UVSOR-III, direct measurements of
undulator spectra around a central wavelength of 355\,nm were obtained 
at a beam current of 0.1\,mA ($1.1\cdot 10^8$ electrons) 
as well as for a single electron using the same setup in both cases.
After passing an aperture (acceptance $\pm 0.3$\,mrad)
and a bandpass filter (wavelength 355\,nm $\pm $ 30\,nm), 
spectra were recorded using a transmission grating and a cooled CCD camera imaging 
radiation diffracted in the first-order.
While background-free 0.1-mA spectra were recorded with an exposure time of 1\,ms, 
accumulating the single-electron spectrum over a time of one hour 
resulted in a spectral distribution on top of a huge background.
However, subtracting a dark CCD image taken over the same time interval yielded
the distributions shown in Fig. 4
for a single undulator (left) and an optical klystron (right) with a small residual background,
which was interpolated between both ends of the spectral range.
It was again confirmed that synchrotron light from a single electron 
has essentially the same spectral properties as 
radiation from a beam with many electrons. 
 
\section{Conclusions}
None of the synchrotron radiation spectra obtained from a single electron show any significant deviation from the spectra of a beam with $10^8$ or more electrons,
which is true for a single undulator as well as the double-pulse interference of an optical klystron.
The usual description of synchrotron radiation within classical electrodynamics
is based on the notion that radiation components from different undulator periods interfere with each other.
The present results demonstrate that a single electron does not emit a photon at a particular magnet within the undulator, which would result in the broad spectrum of a bending magnet,
but the spectral properties are given by delocalized emission 
over the whole magnetic structure.
This is already evident for a single undulator, but was also exemplified for a double-undulator configuration as time-domain analogy of a few-photon double-slit experiment.  
 
The results conform to coherent photon states 
with energies small compared to the electron energy,
which is the case for near-visible light but also for
hard-X-ray synchrotron radiation 
(photon energy of several $10^4$ eV versus an electron energy around $10^9$ eV).    
While being expected by theory,
a direct observation of single-electron spectra is nevertheless intriguing, 
since it differs from conventional quantum optics results in two aspects.
One is that the delocalization of the photon emission extends over one or two undulators, 
macroscopic objects of several meters in length.
The other is that the combination of squared Lorentz factor $\gamma^2$ 
and undulator properties can translate the periodicity of centimeter-sized magnets 
to an extremely wide wavelength range accessible by synchrotron radiation,
from the far-infrared regime to hard X-rays.

A single free relativistic electron in a magnetic field emitting one photon 
is a particularly
simple system calling for further investigations such as 
a study of decoherence effects or delayed-choice experiments.
Another future application could be the implementation of information,
e.g., in electron polarization or the presence/absence of an electron,
or in the properties of the radiating magnetic devices.
Storage rings allow to accommodate a single electron in every radiofrequency (RF) bucket
given by the RF wavelength.
For example, the RF wavelength at DELTA is 0.60 m allowing for 192 buckets along a circumference of 115.2 m.
Phase modulation of the RF wave or transverse island buckets allow to create several 
sub-states within an RF bucket \cite{Jebramcik2022},
which can be controlled using microwave techniques as implemented in bunch-by-bunch feedback systems \cite{Teytelman2020}.
Recent effort under the label of steady-state microbunching (SSMB) aims at replacing 
the bunching RF system by an energy modulation with a laser of, e.g., 
1\,\textmu m wavelength.
While first experimental results are promising \cite{Deng2021,Kruschinski2024},
realizing an SSMB storage ring will require further progress in laser 
and accelerator technology \cite{Li2023}.
Such a storage ring with, for example, 100\,m circumference would have $10^8$ buckets
which could be occupied by single distinguishable electrons.
It may thus be worthwhile to further investigate the properties of 
synchrotron radiation from a single or few relativistic electrons
and techniques to control them.

\section{Methods}
\subsection{Single electrons in storage rings}
The measurements were performed at two different electron storage rings,
DELTA at the TU Dortmund University in Germany 
and UVSOR-III at the Institute of Molecular Science in Okazaki, Japan.
Basic parameters of the respective storage ring and 
the undulators are given in Table 1
and in the text below.

\begin{table}[b]
\caption{Storage ring parameters}\label{tab1}%
\begin{tabular}{@{}lcc@{}}
\toprule
~ & DELTA & UVSOR-III \\
\midrule
Beam energy (MeV)                             & $1500$   & $750$     \\
Lorentz factor                                & $2395$   & $1467$    \\
Circumference (m)                             & $115.2$  & $53.2$    \\
Revolution frequency (MHz)                    & $2.60$   & $5.64$    \\
Radiofrequency (MHz)                          & $500$    & $90.1$    \\
Natural emittance\footnotemark[1] (nm\,rad)   & $16$     & $17$      \\
Bunch length\footnotemark[2] (ps)             & $85$     & $128$     \\
Energy spread\footnotemark[3]                & $7\cdot 10^{-4}$ & $5\cdot 10^{-4}$ \\
Max. multibunch current (A)                   & $0.14$   & $0.30$    \\
Max. single-bunch current (A)                 & $0.02$   & $0.05$    \\
Single-electron current (A)                   & $4.17\cdot 10^{-13}$ & $9.02\cdot 10^{-13}$ \\
Undulator period (mm)                         & $250$    & $88$      \\
Number of periods per undulator               & $4$      & $10$      \\  
\botrule
\end{tabular}
\footnotetext[1]{Product of horizontal beam size and angular divergence.}
\footnotetext[2]{Full width at half maximum in time units.}
\footnotetext[3]{Root mean square value relative to the beam energy.}
\end{table}

In both facilities, the electron beam from a thermionic source is
pre-accelerated by a linear accelerator, boosted to the final energy
in a synchrotron, and then injected into the storage ring.

Since it is difficult to inject exactly one electron, the usual technique
for single-electron experiments is to prepare a beam of low current
and then reduce the beam lifetime to a few minutes 
such that all electrons but one are lost in a controlled way.
For this purpose, a mechanical obstacle, a so-called scraper, is moved close to the beam.
When the beam current is no longer measurable by conventional instruments, 
synchrotron radiation photons from the remaining electrons are counted using 
a photomultiplier tube (PMT) enclosed by a dark box.
The loss of each electron is visible as a discrete step in the PMT count rate, 
as shown in the left part of Fig.\ 5, and
the scraper is retracted when the rate is just one step away from the background
given by dark counts of the PMT and possibly some leakage of light from elsewhere.

The PMT count rate is
$N_\gamma = f_{\rm rev} n_e \varepsilon p_\gamma$,
where the probability of an electron emitting a photon while passing an
undulator $p_\gamma \approx 10^{-2}$ is related to the fine-structure
constant $\alpha \approx 1/137$ \cite{Kim1989},
$n_e$ is the number of electrons circulating with revolution frequency
$f_{\rm rev}$ in the storage ring, and the reduction factor
$\varepsilon $ of $10^{-1}$ to $10^{-2}$ accounts for spectral and spatial constraints
as well as the PMT efficiency.
The red dashed lines in Fig.\ 5 (left) take into account that the 
count rate is reduced by a factor 
$\approx (1-N_\gamma / f_{\rm rev}$),
since the PMT cannot distinguish between a single photon and
multiple hits within its pulse width. 
Thus, the steps are not exactly equal but become larger as $n_e$ decreases.
The assumption of each photon being emitted independently is very well fulfilled, as shown
by comparing the counts measured within 1\,ms with a Poisson distribution -- 
see Fig. 5 (right).

\begin{figure}[t]
\centering
\includegraphics[width=0.8\textwidth]{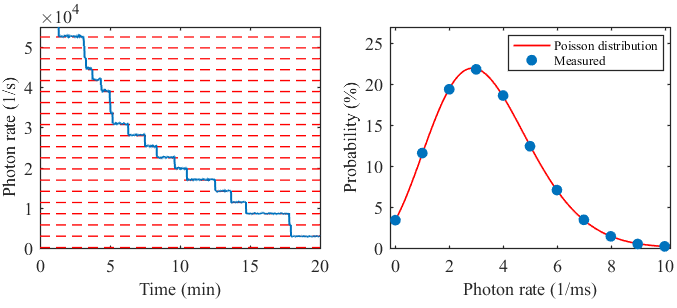}
\caption{Left: Photon rate (blue) while scraping the beam down to a single electron in the DELTA storage ring. Each step (red dashed lines) corresponds to the loss of one electron.
Right: Photon number distribution (blue dots) at UVSOR-III from $10^5$ samples 
of photons counted within an interval of 1\,ms 
together with a fitted Poisson distribution (red line, expectation value 3.38).}
\label{fig5}
\end{figure}

\begin{figure}[t]
\centering
\includegraphics[width=1.0\textwidth]{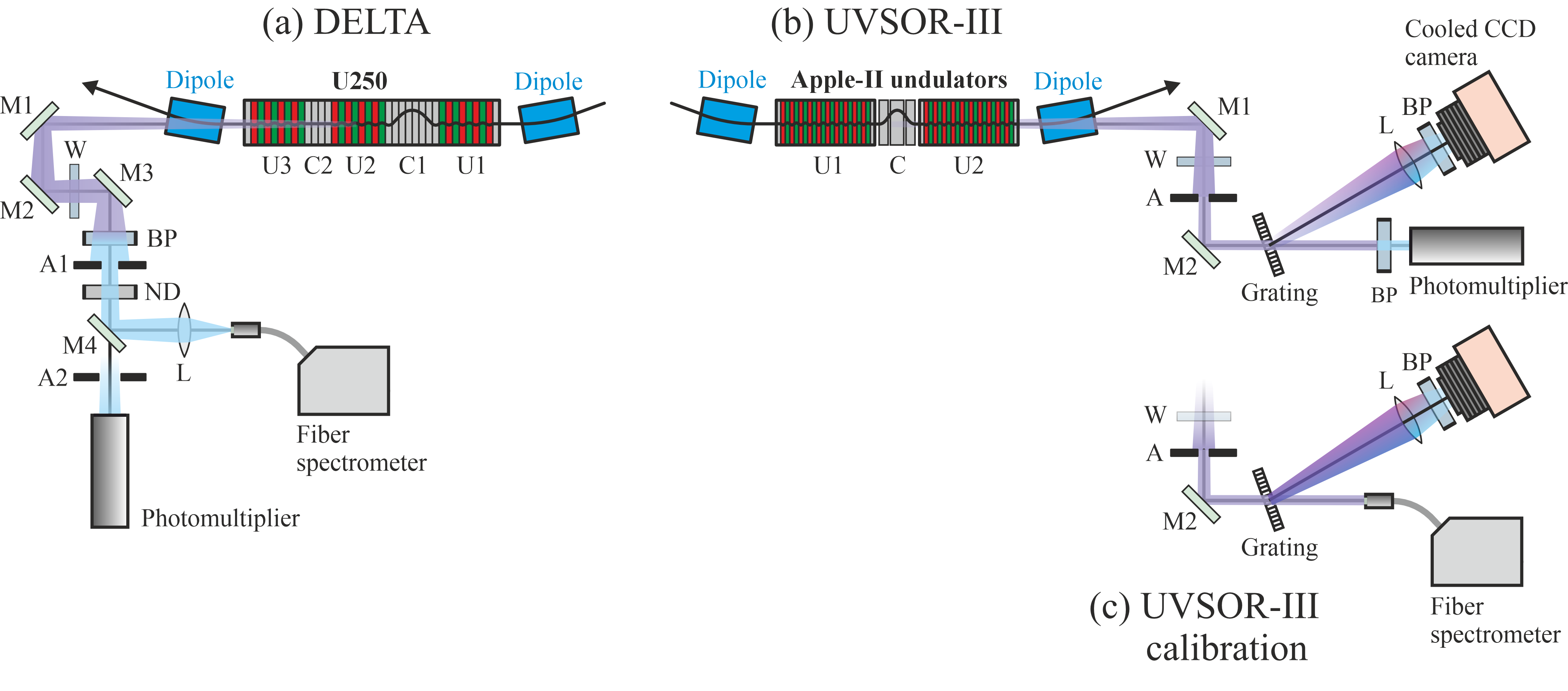}
\caption{Experimental setups at the storage rings DELTA and UVSOR-III.
Synchrotron radiation is generated in magnetic structures including undulators (U) and chicanes (C)
and, to a lesser extent, the adjacent dipole magnets.
The radiation passes a vacuum window (W), mirrors (M), apertures (A), 
bandpass filters (BP) and neutral-density filters (ND).
At DELTA (a), a fiber spectrometer measured radiation from a 2-mA beam while
photons from a single electron were counted using a PMT 
(with ND, M4, and A2 removed).    
At UVSOR-III (b,c), spectra were recorded using a grating and a cooled CCD camera.
The wavelength scale was calibrated by comparing the camera images with a fiber spectrometer.
The PMT was only employed to prepare the single-electron beam.
See text for further details.}
\label{fig6}
\end{figure}

\subsection{Indirect spectral information obtained at DELTA}
At the diagnostics beamline BL 4 of the DELTA storage ring, 
the electromagnetic undulator U250 was employed as single-photon source.
With 19 periods in total, it is presently sub-divided into three undulators of
4 periods each (labeled U1, U2, and U3 in the left part of Fig. 6)
and two magnetic chicanes (C1 and C2) 
to demonstrate echo-enabled harmonic generation (EEHG), 
which was proposed as a seeding scheme for free-electron lasers \cite{Stupakov2009},
but can also be used for short-pulse generation at storage rings \cite{Khan2025}.
With this flexibility, 
single-electron experiments were performed with partial undulator U3 alone
and with two undulators (U1+U2 or U1+U3) in optical-klystron configuration. 

In each measurement, the intensity of radiation passing the vacuum window, a circular aperture (9 mm diameter, 11 m from the U250 center), 
and a bandpass filter (400\,nm $\pm$ 5\,nm) was recorded.
The filter was chosen close to the maximum of the spectral sensitivity of the PMT (Hamamatsu photon counting head H11870-01).
With an electron beam of 2\,mA, sufficient for the orbit correction system to be effective, 
radiation was detected with a fiber spectrometer (Ocean Optics HR2000$+$) comprising a grating and a CCD line camera.
For the case of a single electron, photons were counted with the PMT 
connected to a digital oscilloscope (LeCroy WavePro 804HD),
which recorded $10^8$ samples with a spacing of 10 ns
and delivered the rate of rising signal edges every second.

\begin{figure}[t]
\centering
\includegraphics[width=0.95\textwidth]{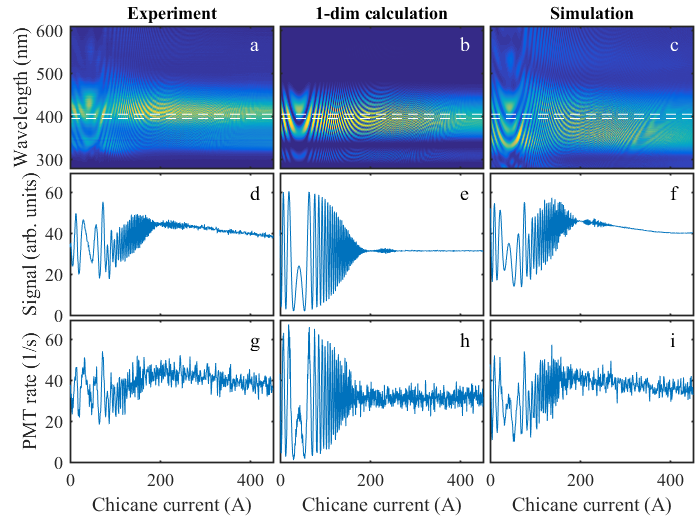}
\caption{Experimental results at DELTA (left), 
analytical calculation (center), and numerical simulation (right). 
Panels (a-c) show spectra without filter under variation of the chicane current, 
(d-f) represent the signal after a bandpass filter (400\,nm $\pm $ 5\,nm) for a 2-mA
electron beam, while (g-i) show the PMT count rate for a single electron.}
\label{fig7}
\end{figure}

In order to obtain spectral information for a comparison between a few-mA beam and a single electron,
the current of the undulator coils was varied, 
thus scanning the $K$ parameter and the fundamental wavelength
of a single undulator and an optical klystron (Fig. 2).
Another measurement in optical-klystron configuration was a scan of the chicane coil current
while keeping the undulator wavelength $\lambda_0$ fixed at 400\,nm (Fig. 3).
This corresponds to scanning the longitudinal dispersion $R_{56}$ and
thus the temporal delay $\Delta t = R_{56}/(2c)$ between pulses from the two
undulators.

The $R_{56}$ value is assumed to be a quadratic function of the coil current 
and was calibrated previously by analyzing optical-klystron spectra.
Since the undulator endpoles are interleaved with the chicanes,
the minimum of $R_{56}$ is not at zero chicane current but at about 42\,A.
The left column of Fig.\ 7 shows measured data: 
Spectra without bandpass filter (top) and summed spectral intensity after inserting the 400-nm filter 
for a 2-mA beam (center) as well as the PMT count rate from a single electron (bottom).

In the central column of Fig.\ 7, 
a one-dimensional on-axis calculation using Eqs.\,(1) to (3)
reproduces some of the experimental features.
Similar to the measured characteristics of the bandpass filter, 
its spectral distribution was modeled in panel (e) as an isosceles triangle with a full width at half maximum of 10\,nm,
which gives rise to the beating of the fringe visibility around a chicane current of 230\,A.
The fluctuations of the PMT count rate are mimicked by adding a random error according to a Poisson distribution
with an expectation value given by the measured rate.

For the right column, synchrotron radiation within the experimental acceptance angle
was simulated using the code SPECTRA \cite{Tanaka2001}.
The magnetic field of undulators, chicane, and dipole magnets were 
computed with the magnetostatic field solver of CST Microwave Studio \cite{CSTStudio}.
Again, a bandpass filter and Poisson statistics was applied to calculate the 2-mA
and single-electron signals as functions of chicane current.
The simulation shows that the results
are significantly modified by the finite aperture as well as
by radiation from the chicane magnets and the dipoles enclosing the straight section.
Interference from multiple dipole edges \cite{Geloni2009} adds spectral features that depend on
the chicane settings.
  

For a detailed comparison between measurement and simulation, 
some experimental shortcomings must be taken into account,
most of them reducing the observed fringe visibility. 
The solid angle accepted by the aperture A1 in
Fig. 6 (left) may be slightly off-center and could have changed over time
since the orbit correction system cannot operate with a single electron.
Furthermore, the optics of the electron beam and the
permeability of the magnetic material
may differ from the assumptions. 
Despite applying a cycling procedure, 
the relationship between chicane current and resulting $R_{56}$ value
is never identical between two measurements.
Within these uncertainties,
the numerical simulation is consistent with the experimental data,
and no significant discrepancy between the few-mA and single-electron cases can be claimed.

\subsection{Direct measurement of undulator spectra at UVSOR-III}
At UVSOR-III, single-electron experiments were performed at beamline BL1U \cite{Katoh2024,Kaneyasu2025}
using two identical permanent-magnet APPLE-II undulators with 10 periods each (labeled U1 and U2 in right part of Fig. 6) 
and a three-magnet chicane (C) between them.
These undulators are capable of producing any degree of elliptical polarization,
but were set to linear horizontal polarization and a central wavelength of 355\,nm 
with either one undulator alone or both tuned to the same wavelength in an optical-klystron configuration.

\begin{figure}[t]
\centering
\includegraphics[width=0.7\textwidth]{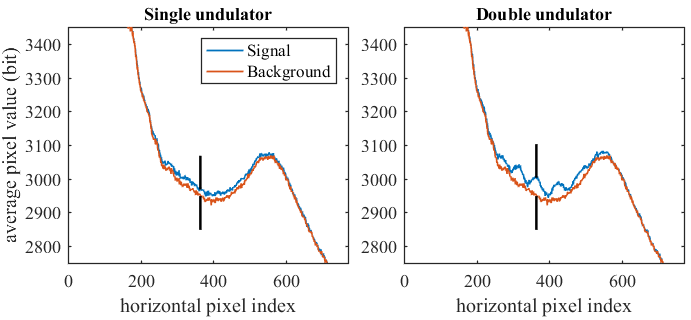}
\caption{Vertical projections of single-electron CCD images for one undulator (left) and two consecutive undulators (right) at UVSOR-III. 
Blue: Average of vertical CCD pixel values as function of the horizontal pixel index (representing the wavelength axis) for a single electron.
Red: Background from a dark image with the same exposure time of 3600\,s.
At the marker positions, the signal-to-background ratio is 0.007 
and 0.018 for one and two undulators, respectively.}
\label{fig8}
\end{figure}

After a 90-degree deflection by an Al mirror in ultrahigh vacuum, the radiation was extracted through a sapphire vacuum window into air, passed a circular aperture (5\,mm diameter, 8\,m from the center of the double-undulator), 
and was directed onto a transmission grating (830 grooves/mm).
In order to prepare a single-electron beam as described above, zeroth-order radiation
from the grating was observed using a PMT.
First-order light was imaged by a cooled CCD camera (BITRAN BS-40L), where
the horizontal coordinate represented the wavelength scale.
A bandpass filter (wavelength 355\,nm $\pm $ 60\,nm) was attached to the camera to reduce ambient background radiation.
Spectral measurements were conducted for a beam of 0.1-mA with an exposure time of 1\,ms. 
In the case of a single electron, the exposure time of the camera was set to its maximum 
(3600\,s).
Despite an unfavorable signal-to-background ratio
(see Fig.\ 8), subtracting a dark image
with the same exposure time shows a clear 
undulator spectrum and the expected interference pattern in the case of two undulators.
Taking the average of the vertical pixel values, single-electron spectra were obtained  -- see gray and blue dots in Fig. 4.
The surprisingly low residual background was linearly interpolated
between spectral regions suppressed by the bandpass filter.
To calibrate the wavelength scale, the undulator $K$ parameter was varied to shift the central wavelength $\lambda_0$ in several steps.
At each setting, a camera image was recorded together with zeroth-order light captured by a fiber spectrometer (Ocean Optics Flame-T),
and a linear fit was performed between the wavelength at the spectral maximum and the respective camera pixel.


\bmhead{Acknowledgements}

The measurements at DELTA were funded by the Federal Ministry of Research, Technology and Space (grant 05K22PE4) 
and by the State North Rhine-Westphalia. 
The work at UVSOR-III was supported by the Japan Society for the Promotion of Science 
(KAKENHI grants 22K18705 and 25H00623) 
and by the Institute of Molecular Science
(grants 22IMS6868, 23IMS6610, and 24IMS6809).
The authors would like to thank the teams operating their respective
synchrotron radiation facility.

\section*{Competing interests}
The authors declare no competing interests. 






\bibliography{single_electron_bib}

@PREAMBLE{
 "\providecommand{\noopsort}[1]{}" 
 # "\providecommand{\singleletter}[1]{#1}%" 
}

@inbook{Teytelman2020,
   author = {Teytelman, D.},
   title = {Coupled-bunch instabilities in storage rings and feedback},
   booktitle = {Synchrotron Light Sources and Free-Electron Lasers {(2nd Ed.)}},
   editor = {Jaeschke, E. J. and Khan, S. and Schneider, J. R. and Hastings, J. B.},
   series = {Springer Reference},
   publisher = {Springer},
   address = {Cham, Switzerland},
   year = {2020},
   pages = {683},
   doi = {10.1007/978-3-319-14394-1}
}

@BOOK{Fox2006,
   author = "{Fox, M.}",
   title = "Quantum Optics",
   publisher = "Oxford University Press",
   address = "Oxford, UK",
   year = "2006",
   note = ""
}

@BOOK{Bachor2019,
   author = "{Bachor, H.-A., and Ralph, T. C.}",
   title = "A Guide to Experiments in Quantum Optics",
   publisher = "Wiley-VCH",
   address = "Weinheim, Germany",
   year = "2019",
   note = ""
}

@BOOK{Ou2017,
   author = "{Ou, Z.-Y. J.}",
   title = "Quantum Optics for Experimentalists",
   publisher = "World Scientific",
   address = "Singapore",
   year = "1917",
   note = ""
}

@BOOK{Wiedemann2010,
   author = "{Wiedemann, H.}",
   title = "Synchrotron Radiation",
   publisher = "Springer",
   address = "Berlin, Germany",
   year = "2010",
   note = ""
}

@BOOK{Hofmann2007,
   author = "{Hofmann, A.}",
   title = "The Physics of Synchrotron Radiation",
   publisher = "World Scientific",
   address = "Singapore",
   year = "1917",
   note = ""
}

@BOOK{Jackson1999,
   author = "{Jackson, J. D.}",
   title = "{Classical Electrodynamics}",
   publisher = "John Wiley \& Sons",
   address = "Hoboken, NJ, USA",
   year = "1999",
   note = ""
}

@ARTICLE{Young1804, 
   author = "{Young, T.}", 
   title = "The {Bakerian} lecture. {Experiments} and calculation relative to physical optics", 
   journal = "Philosophical Transactions of the Royal Society of London",
   volume = "94",
   pages = "1--16",
   year = "1804",
   doi = "10.1098/rstl.1804.0001"
}

@ARTICLE{Zeilinger1988,
   author = "{Zeilinger, A., Gähler, R., Shull, C. G., Treimer, W., and Mampe W}", 
   title = "Single- and double-slit diffraction of neutrons", 
   journal = "Review of Modern Physics",
   volume = "60",
   pages = "1067--1073",
   year = "1988",
   doi = "10.1103/RevModPhys.60.1067"
}

@ARTICLE{Joensson1961,
   author = "{Jönsson, C.}", 
   title = "{Elektroneninterferenzen an mehreren künstlich hergestellten Feinspalten}", 
   journal = "Zeitschrift für Physik",
   volume = "161",
   pages = "454--474",
   year = "1961",
   doi = "10.1007/BF01342460"
}

@ARTICLE{Taylor1909,
   author = "{Taylor, G. I.}", 
   title = " Interference fringes with feeble light", 
   journal = "Proceedings of the Cambridge Philosophical Society",
   volume = "15",
   pages = "114--115",
   year = "1909"
}

@article{Brezger2002,
  title = {Matter-Wave Interferometer for Large Molecules},
  author = {Brezger, Bj\"orn and Hackerm\"uller, Lucia and Uttenthaler, Stefan and Petschinka, Julia and Arndt, Markus and Zeilinger, Anton},
  journal = {Phys. Rev. Lett.},
  volume = {88},
  issue = {10},
  pages = {100404},
  year = {2002},
  doi = {10.1103/PhysRevLett.88.100404}
}

@article{Kruschinski2024,
  title = {Confirming the theoretical foundation of steady-state microbunching},
  author = {A. Kruschinski and X. Deng and J. Feikes and A. Hoehl and R. Klein and J. Li and M. Ries and A. Chao},
  journal = {Commun. Phys.},
  volume = {7},
  pages = {160},
  year = {2024},
  doi = {10.1038/s42005-024-01657-y}
}

@article{Deng2021,
  title = {Experimental demonstration of the mechanism of steady-state microbunching},
  author = {X. Deng and A. Chao and J. Feikes and A. Hoehl and W. Huang and R. Klein and A. Kruschinski and J. Li and A. Matveenko and Y. Petenev and M. Ries and C. Tang and L. Yan},
  journal = {Nature},
  volume = {590},
  pages = {576--579},
  year = {2021},
  doi = {10.1038/s41586-021-03203-0}
}

@article{Li2023,
  title = {Generalized longitudinal strong focusing in a steady-state microbunching storage ring},
  author = {Z. Li and X. Deng and Zh. Pan and Ch. Tang and A. Chao},
  journal = {Phys. Rev. Accel Beams},
  volume = {16},
  pages = {110701},
  year = {2023},
  doi = {10.1103/PhysRevAccelBeams.26.110701}
}

@ARTICLE{Pinayev1994, 
   author = "{Pinayev, I. V.,  Popik, V. M., Shaftan, T. V., Sokolov, A. S., Vinokurov, N. A., and Vorobyov, P. V.} ", 
   title = "Experiments with undulator radiation of a single electron", 
   journal = "Nuclear Instruments and Methods in Physics Research A",
   volume = "341",
   pages = "17--20",
   year = "1994",
   doi = "10.1016/0168-9002(94)90308-5"
}

@ARTICLE{Aleshaev1995, 
   author = "{Aleshaev, A. N., Pinayev, I. V., Popik, V. M., Serednyakov, S. S., Shaftan, T. V., Sokolov, A. S., Vinokurov, N. A., and Vorobyov, P. V.}", 
   title = "A study of the influence of the stochasitic process on the synchrotron oscillations of a single electron using undulator radiation", 
   journal = "Nuclear Instruments and Methods in Physics Research A",
   volume = "359",
   pages = "80--84",
   year = "1995",
   doi = "10.1016/0168-9002(96)88028-4"
}

@ARTICLE{Pinayev1996, 
   author = "{Pinayev, I. V., Popik, V. M., Salikova, T. V., Shaftan, T. V., Sokolov, A. S., Vinokurov, N. A., and Vorobyov, P. V.}", 
   title = "{A study of the influence of the stochasitic process on the synchrotron oscillations of a single electron circulated in the VEPP-3 storage ring}", 
   journal = "Nuclear Instruments and Methods in Physics Research A",
   volume = "375",
   pages = "71--73",
   year = "1996",
   doi = "10.1016/0168-9002(95)01350-4"
}

@ARTICLE{Riehle1988, 
   author = "{Riehle, F., Bernstorff, S., Fr\"ohling, R., and Wolf, F.}", 
   title = "{Determination of electron currents below 1 nA in the storage ring BESSY by measurement of the synchrotron radiation of single electrons}", 
   journal = "Nuclear Instruments and Methods in Physics Research A",
   volume = "268",
   pages = "262--269",
   year = "1988",
   doi = "10.1016/0168-9002(88)90616-X"
}

@ARTICLE{Brandt2007, 
   author = "{Brandt, G., Eden, J., Fliegauf, R., Gottwald, A., Hoehl, A., Klein, R., M\"uller, R., Richter, M., Scholze, F., Thornagel, R., Ulm, G., B\"urkmann, K., Rahn, J., and W\"ustefeld, G.}", 
   title = "The Metrology Light Source – {The} new dedicated electron storage ring of {PTB}", 
   journal = "Nuclear Instruments and Methods in Physics Research B",
   volume = "258",
   pages = "445--452",
   year = "2007",
   doi = "10.1016/j.nimb.2007.02.076"
}

@ARTICLE{Klein2010, 
   author = "{Klein, R., Thornagel, R., and Ulm, G.}", 
   title = "From single photons to milliwatt radiant power -- electron storage rings as radiation sources with a high dynamic range", 
   journal = "Metrologia",
   volume = "47",
   pages = "R33",
   year = "2010",
   doi = "10.1103/PhysRevSTAB.11.110701"
}

@ARTICLE{Romanov2021, 
   author = "Romanov, A. and Santucci, J. and Stancari, G. and Valishev, A. and Kuklev, N.", 
   title = "Experimental 3-dimensional tracking of the dynamics of a single electron in the {Fermilab Integrable Optics Test Accelerator (IOTA)}", 
   journal = "Journal of Instrumentation",
   volume = "16",
   pages = "P12009",
   year = "2021",
   doi = "10.1088/1748-0221/16/12/P12009"
}

@ARTICLE{Lobach2020, 
   author = "{Lobach, I., Lebedev, V., Nagaitsev, S. , Romanov, A., Stancari, G., Valishev, A., Halavanau, A., Huang., Zh., and Kim K.-J.}", 
   title = "Statistical properties of spontaneous synchrotron radiation with arbitrary degree of coherence", 
   journal = "Physical Review Accelerators and Beams",
   volume = "23",
   pages = "090703",
   year = "2020",
   doi = "10.1103/PhysRevAccelBeams.23.090703"
}

@ARTICLE{Lobach2022, 
   author = "{Lobach, I., Nagaitsev, S., Romanov, and Stancari, G.}", 
   title = "Single electron in a storage ring: a probe into the fundamental properties of synchrotron radiation and a powerful diagnostic tool", 
   journal = "Journal of Instrumentation",
   volume = "17",
   pages = "P02014",
   year = "2022",
   doi = "10.1088/1748-0221/17/02/P02014"
}

@ARTICLE{Bonolis2018, 
   author = "{Bonolis, L. and Pancheri, G.}", 
   title = "{Bruno Touschek and AdA: from Frascati to Orsay}", 
   journal = "INFN, Laboratori Nationali di Frascati",
   volume = "INFN - 18-05/LNF",
   pages = "",
   year = "2018",
   doi = "10.48550/arXiv.1805.09434"
}

@INPROCEEDINGS{Stancari2024, 
   author = "{Stancari, G., Jarvis, J., Romanov, A. Shemyakin, A., Valishev, A., Lobach, I., and Nagaitsev, H.}",
   title = "{Undulator radiation of single electrons: Coherence length and quantum-optical properties}",
   booktitle = "{Proceedings of the 15th International Particle Accelerator Conference (IPAC'24), 
                     Nashville, TN, USA}",
   year = "2024",
   pages = "300--303",
   publisher = "{JACoW} Publishing",
   address = "Geneva, Switzerland",
   doi = "10.18429/JACoW-IPAC2024-MOPG06"
}

@INPROCEEDINGS{Khan2024, 
   author = "{Khan, S., Büsing, B., Mai, C., Radha Krishnan, A. Usfoor, Z., Vijayan, V., and Salah, W.}",
   title = "{Single-electron experiments at the DELTA storage ring}",
   booktitle = "{Proceedings of the 15th International Particle Accelerator Conference (IPAC'24), 
                     Nashville, TN, USA}",
   year = "2024",
   pages = "1358--1360",
   publisher = "{JACoW} Publishing", 
   address = {Geneva, Switzerland},
   doi = "10.18429/JACoW-IPAC2024-TUPG51"
}

@INPROCEEDINGS{Khan2025, 
   author = "{Khan, S., Büsing, B., Mai, C., Radha Krishnan, A. Usfoor, Z., Vijayan, V., and Salah, W.}",
   title = "Echo-enabled harmonic generation at the {DELTA} storage ring",
   booktitle = "{Proceedings of the 16th International Particle Accelerator Conference (IPAC'25), 
                     Taipei, Taiwan}",
   year = "2025",
   pages = "87--90",
   publisher = "{JACoW} Publishing",
   address = {Geneva, Switzerland},
   doi = "10.18429/JACoW-IPAC2025-MOPB007"
}

@INPROCEEDINGS{Asai2025, 
   author = "{Asai, Y., Shimada, M., Miyauchi, H., Katoh, M., and Kaneyasu, T.}",
   title = "{Single electron storage at {UVSOR-III} electron storage ring}",
   booktitle = "{Proceedings of the 16th International Particle Accelerator Conference (IPAC'25), 
                     Taipei, Taiwan}",
   year = "2025",
   pages = "745--747",
   publisher = "{JACoW} Publishing",
   address = "Geneva, Switzerland",
   doi = "10.18429/JACoW-IPAC25-MOPS056"
}

@ARTICLE{Elleaume1983, 
   author = "Elleaume, P.", 
   title = "Optical klystrons", 
   journal = "Journal de Physique Colloques",
   volume = "44",
   pages = "C1-333--352",
   year = "1983",
   doi = "10.1051/jphyscol:1983127"
}

@ARTICLE{Kornyukhin1985, 
   author = "{Kornyukhin, G. A., Kulipanov, G. N.,  Litvinenko, V. N., Mesentsev, N. A., Skrinsky, A. N., Vinokurov, N. A., and Voblyi, P. D.}", 
   title = "Status of the {INP} optical klystron", 
   journal = "Nuclear Instruments and Methods in Physics Research A",
   volume = "237",
   pages = "281--288",
   year = "1985",
   doi = "10.1016/0168-9002(85)90362-6"
}

@ARTICLE{Geloni2009, 
   author = "{Geloni, G., Kocharyan, V., Saldin E., Schneidmiller, E., and Yurkov, M.}", 
   title = "{Theory of edge radiation. Part I: Foundations and basic applications}", 
   journal = "Nuclear Instruments and Methods in Physics Research A",
   volume = "605",
   pages = "409--429",
   year = "2009",
   doi = "10.1016/j.nima.2009.03.240"
}

@ARTICLE{Kim1989, 
   author = "{Kim, K.-J.}", 
   title = "Characteristics of synchrotron radiation", 
   journal = "AIP Conference Proceedings",
   volume = "184",
   pages = "565--632",
   year = "1989",
   doi = "10.1063/1.38046"
}

@ARTICLE{Tanaka2001, 
   author = "{Tanaka, T. and Kitamura, H.}", 
   title = "{SPECTRA:} a synchrotron radiation calculation code", 
   journal = "Journal of Synchrotron Radiation",
   volume = "8",
   pages = "1221--1228",
   year = "2001",
   doi = "10.1107/S090904950101425X"
}

@misc{CSTStudio,
   title = "{CST Studio Suite, Dassault System\`es}",
   note = "https://www.3ds.com/"
}

@ARTICLE{Stupakov2009,
   author = "Stupakov, G.",
   title = "{Using the beam-echo effect for generation of short-wavelength radiation}", 
   journal = "Physical Review Letters",
   volume = "102",
   pages = "074801",
   year = "2009",
   doi = "10.1103/PhysRevLett.102.074801"
}

@ARTICLE{Jebramcik2022, 
   author = "{Jebramcik, M., Khan, S., and Helml, W.}", 
   title = "Steady-state solutions of split beams in electron storage rings", 
   journal = "Scientific Reports",
   volume = "12",
   pages = "18383",
   year = "2022",
   doi = "10.1038/s41598-022-22857-y"
}

@ARTICLE{Katoh2024, 
   author = "M. Katoh and H. Ota and J. Yamazaki and K. Hayashi and Y. Okano and E. Salehi and Y. Taira and A. Mano and M. Fujimoto and Y. Takashima and M. Hosaka and F. Sakamoto and T. Kaneyasu and H. Zen", 
   title = "Light Source Developments at {UVSOR BL1U}", 
   journal = "Journal of Physics: Conference Series",
   volume = "2687",
   pages = "032005",
   year = "2024",
   doi = "10.1088/1742-6596/2687/3/032005"
}

@ARTICLE{Kaneyasu2025, 
   author = "T. Kaneyasu and Y. Hikosaka and S. Wada and H. Ota and H. Iwayama and K. Shimizu and M. Fujimoto and M. Katoh", 
   title = "Attosecond Interferometry Experiments at the Tandem Undulator Beamline {BL1U} of {UVSOR-III} Synchrotron", 
   journal = "Journal of Physics: Conference Series",
   volume = "3010",
   pages = "012086",
   year = "2025",
   doi = "10.1088/1742-6596/3010/1/012086"
}

\end{document}